\documentclass[10pt,a4paper]{article}

\usepackage{arxiv}

\usepackage[utf8]{inputenc}
\usepackage[T1]{fontenc}
\usepackage{amsmath,amssymb,amsfonts}
\usepackage{graphicx}
\usepackage{booktabs}
\usepackage{multirow}
\usepackage{bm}
\usepackage{float}
\usepackage{hyperref}
\usepackage{cleveref}
\usepackage{siunitx}
\usepackage{subcaption}
\usepackage{enumitem}
\usepackage{natbib}

\graphicspath{{figures/}}

\title{Nonlinear Flexibility Effects on Flight Dynamics\\of High-Aspect-Ratio Wings}

\author{
  Nikolaos D.~Tantaroudas\thanks{Corresponding author. Senior Researcher, ICCS.} \\
  Institute of Communications and Computer Systems (ICCS)\\
  9 Iroon Politechniou Street, Zografou, Athens 15773, Greece\\
  \texttt{nikolaos.tantaroudas@iccs.gr} \\
  \And
  Ilias Karachalios \\
  National Technical University of Athens\\
  Zografou, Athens 15780, Greece\\
}

\date{}

\begin{document}

\maketitle

\begin{abstract}
This paper investigates the effects of geometric nonlinearity and structural flexibility on the flight dynamics of high-aspect-ratio wings representative of high-altitude long-endurance (HALE) aircraft configurations. A coupled aeroelastic--flight dynamic framework is developed, combining a geometrically-exact beam formulation for the structure, unsteady two-dimensional strip theory for the aerodynamics, and quaternion-based rigid-body equations for the flight dynamics. The three subsystems are monolithically coupled through consistent load and motion transfer at each time step. A systematic parametric study is conducted by varying the wing stiffness across several orders of magnitude, spanning from nearly rigid to very flexible configurations. The study reveals that increasing flexibility fundamentally alters trim conditions, flutter boundaries, and dynamic gust response. In particular, large static deformations create an effective dihedral that modifies the lift direction and necessitates higher trim angles of attack. The phugoid mode is shown to destabilise at high flexibility levels, consistent with findings in the literature. Flutter speed degradation is quantified as a function of the stiffness parameter, showing significant reductions for very flexible configurations when the pre-stressed equilibrium is correctly accounted for. The framework is validated against published HALE aircraft benchmarks, demonstrating good agreement in natural frequencies, flutter speeds, and nonlinear static deflections. Results provide quantitative guidance on when linear analysis is acceptable and when fully coupled nonlinear tools become essential.
\end{abstract}


\section{Introduction}
\label{sec:introduction}

The design of modern high-altitude long-endurance (HALE) and solar-powered unmanned aerial vehicles (UAVs) has pushed the boundaries of conventional aeroelastic and flight dynamic analysis. Aircraft such as NASA's Helios~\cite{noll2004investigation}, the Airbus Zephyr, and various solar-powered platforms feature high-aspect-ratio wings with extremely low structural weight fractions, resulting in wing-tip deflections that can exceed 25\% of the semi-span during normal flight~\cite{patil2001nonlinear,tantaroudas2017bookchapter}. At such deformation levels, the conventional separation of structural, aerodynamic, and flight dynamic analyses, each conducted independently with linearised assumptions, becomes inadequate.

The catastrophic loss of the Helios prototype in June 2003~\cite{noll2004investigation} underscored the need for integrated aeroelastic--flight dynamic tools capable of capturing the nonlinear coupling between large structural deformations and the rigid-body dynamics of very flexible aircraft (VFA). Post-accident analysis identified an unexpected interaction between structural flexibility, aerodynamic forces, and the flight control system as a contributing factor, highlighting that the coupled physics cannot be reliably predicted by standard methods. The mishap report specifically noted that ``the existing analytical tools used during development of the Helios configuration were not adequate to predict the vehicle's stability characteristics under the atmospheric conditions encountered'', a finding that motivated the present and related research programmes.

Significant progress has been made in developing computational frameworks for VFA analysis. Patil, Hodges, and Cesnik~\cite{patil2001nonlinear,patil1999nonlinear} pioneered the use of geometrically-exact beam formulations coupled with finite-state aerodynamics for HALE aircraft, demonstrating that flexibility has a profound effect on flight dynamic stability, particularly the phugoid mode. In a seminal study, Patil and Hodges~\cite{patil2001flight} showed that the phugoid mode of a representative HALE configuration becomes unstable when wing flexibility is accounted for, a result that was entirely unpredicted by conventional rigid-body flight dynamic analysis. Hodges~\cite{hodges2003geometrically,hodges2006nonlinear} developed the variational-asymptotic beam sectional analysis (VABS) and the geometrically-exact intrinsic beam theory that underpins much of the subsequent literature. Cesnik and Brown~\cite{cesnik2002modeling} extended these ideas to the active aeroelastic tailoring of HALE wings, while Shearer and Cesnik~\cite{shearer2007nonlinear} developed a comprehensive nonlinear flight dynamics simulation capability for very flexible aircraft.

The aerodynamic modelling fidelity has also evolved. Murua et al.~\cite{murua2012applications,murua2012coupled} coupled an unsteady vortex-lattice method (UVLM) with a geometrically-nonlinear beam model, providing improved predictions of the unsteady wake effects on flexible aircraft stability. Their work demonstrated that three-dimensional aerodynamic effects can alter flutter predictions by 5--10\% compared to strip-theory-based approaches, depending on the aspect ratio and reduced frequency range. Hesse and Palacios~\cite{hesse2014reduced,hesse2016consistent} developed reduced-order models for the consistent coupling of nonlinear beams with unsteady aerodynamics, and demonstrated the importance of maintaining geometric consistency between the structural and aerodynamic reference frames. Their work highlighted that inconsistent linearisation, linearising the structure about the deformed state but using undeformed aerodynamic geometry, can introduce errors of comparable magnitude to those caused by neglecting nonlinearity altogether.

Tantaroudas and Da Ronch~\cite{tantaroudas2017bookchapter,tantaroudas2015nonlinear} developed nonlinear reduced-order aeroservoelastic models for very flexible aircraft, enabling efficient control design on systems with thousands of degrees of freedom. This work established a framework for model order reduction that preserves the essential nonlinear coupling between structural and flight dynamic modes, even when the structural model is projected onto a small number of modal coordinates. Tantaroudas et al.~\cite{tantaroudas2015scitech} further extended this approach to control design of flexible free-flying aircraft, demonstrating that the reduced-order models can be used directly in model-based control synthesis. Da Ronch et al.~\cite{daronch2014scitech} systematically assessed the impact of aerodynamic modelling fidelity on manoeuvring aircraft, comparing strip theory, doublet-lattice, UVLM, and RANS-based approaches for a range of flexible configurations. Da Ronch et al.~\cite{daronch2013gust,daronch2013control} further investigated model reduction techniques for both linear and nonlinear gust loads analysis and control applications, establishing the accuracy requirements for reduced-order models in certification-level gust encounter simulations. A self-contained derivation of the NMOR formulation, including third-order Taylor expansion terms and systematic eigenvector selection criteria, is presented in the companion paper~\cite{Tantaroudas2026nmor}. The application of the same ROM framework to rapid worst-case gust identification and $\mathcal{H}_\infty$ robust control for gust load alleviation of geometrically nonlinear flexible aircraft are respectively demonstrated in~\cite{Tantaroudas2026gust} and~\cite{Tantaroudas2026hinf}.

More recent work has explored higher-fidelity aerodynamic coupling and advanced reduced-order modelling. Wang et al.~\cite{wang2016nonlinear} investigated CFD/CSD coupling for high-aspect-ratio flexible wings, demonstrating the importance of viscous effects at high angles of attack but also confirming that inviscid methods provide adequate accuracy at moderate incidence. Deskos et al.~\cite{deskos2020review} provided a comprehensive review of computational methods for flexible aircraft aeroelasticity, highlighting the trade-offs between fidelity, computational cost, and suitability for parametric design studies. Su and Cesnik~\cite{sucesnik2010} examined the nonlinear aeroelasticity of blended-wing-body configurations, extending the established beam-based methodologies to unconventional planforms. Artola et al.~\cite{Artola2021} demonstrated the viability of aeroelastic control and state estimation using a minimal nonlinear modal description combined with moving-horizon estimation and model-predictive control strategies. Goizueta et al.~\cite{Goizueta2022} introduced adaptive sampling techniques for interpolation of parametric reduced-order aeroelastic models across multiple flight conditions and design points. Riso and Cesnik~\cite{Riso2023} conducted a systematic assessment of the accuracy of low-order modelling approaches for aeroelastic predictions of very flexible wings, confirming that beam-based ROM frameworks are well suited to the large-deformation regime. Del Carr\'{e} and Palacios~\cite{DelCarre2019} developed efficient time-domain simulation techniques for nonlinear aeroelastic problems. A comprehensive treatment of the coupled flight mechanics, aeroelasticity, and control of flexible aircraft has recently been provided in the monograph by Palacios and Cesnik~\cite{PalaciosCesnik2023}.

The development of active control strategies for very flexible aircraft has progressed in parallel with the modelling advances. Tantaroudas et al.~\cite{tantaroudas2014aviation} developed an adaptive aeroelastic control approach using nonlinear reduced-order models for gust load alleviation, demonstrating robustness to variations in flight condition and structural parameters. Da Ronch et al.~\cite{daronch2014flutter} developed a nonlinear controller for flutter suppression that was validated through wind tunnel testing, bridging the gap between numerical prediction and experimental demonstration. Papatheou et al.~\cite{papatheou2013ifasd} and Fichera et al.~\cite{fichera2014isma} further contributed experimental investigations of active flutter control and nonlinear dynamic behaviour of aerofoil sections. More recently, Tantaroudas et al.~\cite{Tantaroudas2026mrac} developed a model reference adaptive control (MRAC) framework grounded in Lyapunov stability theory for gust load alleviation of nonlinear aeroelastic systems, demonstrating significant wing-tip deflection reductions on both discrete and stochastic gust encounters. In a related theoretical contribution, Tantaroudas~\cite{Tantaroudas2026ballbeam} established the minimum number of control laws required for nonlinear systems exhibiting input-output linearisation singularities, providing foundational results applicable to the control of nonlinear aeroelastic systems.

Despite these advances, a systematic parametric investigation of how the degree of structural flexibility affects the full spectrum of flight dynamic behaviour, from trim through stability to dynamic gust response, remains limited. Most published studies focus on a single aircraft configuration or a narrow range of stiffness values. The present work addresses this gap by introducing a non-dimensional stiffness parameter $\sigma$ and conducting a comprehensive study across four orders of magnitude of wing stiffness. The objectives are fourfold: first, to validate the coupled framework against established HALE benchmarks, including natural frequencies, static deflections, and flutter speeds; second, to quantify how flexibility alters trim conditions, with particular emphasis on the lift-vector rotation mechanism; third, to determine the flutter speed degradation as a function of $\sigma$ and to identify the physical mechanisms responsible; and fourth, to characterise the dynamic gust response across the full range of flexibility and to provide practical guidance on when nonlinear, fully-coupled analysis becomes necessary.

The remainder of this paper is organised as follows. Section~\ref{sec:model} describes the coupled aeroelastic--flight dynamic model, including the structural, aerodynamic, and rigid-body subsystems and their monolithic coupling. Section~\ref{sec:parameterStudy} defines the flexibility parameter study and the test aircraft configuration. Section~\ref{sec:results} presents the validation and parametric results. Section~\ref{sec:discussion} discusses the practical implications, and Section~\ref{sec:conclusions} summarises the main conclusions.

\section{Coupled Aeroelastic--Flight Dynamic Model}
\label{sec:model}

\begin{figure}[htbp]
  \centering
  \includegraphics[width=0.85\textwidth]{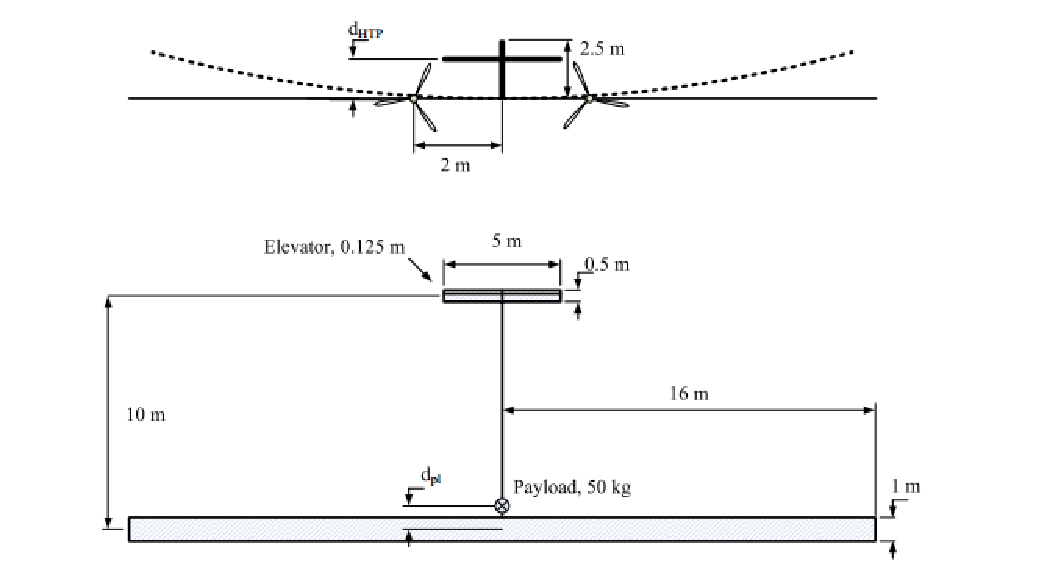}
  \caption{HALE aircraft geometry and coordinate frame definitions: inertial frame $\mathcal{A}$, body-fixed frame $\mathcal{B}$, and local beam cross-sectional frame $\mathcal{E}$. The wing undergoes large bending and torsion described by the geometrically-exact beam formulation.}
  \label{fig:haleGeometry}
\end{figure}

The coupled framework consists of three subsystems: a geometrically-exact beam model for the structure (\S\ref{subsec:structural}), an unsteady strip-theory aerodynamic model (\S\ref{subsec:aero}), and a quaternion-based six-degree-of-freedom (6-DOF) flight dynamic model (\S\ref{subsec:flightDyn}). These are monolithically coupled through consistent transformations of loads and kinematics (\S\ref{subsec:coupling}). The overall architecture is illustrated schematically in \Cref{fig:haleGeometry}, which shows the HALE aircraft geometry and the coordinate frames used throughout the formulation.

\subsection{Structural Model: Geometrically-Exact Beam Theory}
\label{subsec:structural}

The wing structure is modelled as a one-dimensional beam undergoing arbitrarily large displacements and rotations, using the geometrically-exact formulation~\cite{hodges2003geometrically,palacios2010assessment}. This approach retains the full geometric nonlinearity without approximation, making it suitable for the large deformations encountered in HALE aircraft wings. Each cross-section of the beam is characterised by six degrees of freedom: three translations $(u_1, u_2, u_3)$ and three rotations $(\theta_1, \theta_2, \theta_3)$ with respect to a body-fixed reference frame $\mathcal{B}$.

The position vector of a point on the deformed beam reference line is
\begin{equation}
  \bm{r}(s,t) = \bm{r}_0(s) + \bm{u}(s,t),
  \label{eq:position}
\end{equation}
where $\bm{r}_0(s)$ is the undeformed position, $\bm{u}(s,t)$ is the displacement vector, and $s$ denotes the curvilinear coordinate along the beam span. The deformed beam reference line is assumed to be inextensible in the initial formulation, although the extensional degree of freedom is retained in the stiffness matrix for generality.

The orientation of each cross-section is described by a rotation tensor $\bm{C}^{Ba}(s,t)$ that maps vectors from the inertial frame $\mathcal{A}$ to the local beam frame $\mathcal{B}$. This tensor is parameterised using the Cartesian rotation vector $\bm{\psi}$, such that
\begin{equation}
  \bm{C}^{Ba} = \bm{I} + \frac{\sin\|\bm{\psi}\|}{\|\bm{\psi}\|}\widetilde{\bm{\psi}} + \frac{1 - \cos\|\bm{\psi}\|}{\|\bm{\psi}\|^2}\widetilde{\bm{\psi}}^2,
  \label{eq:rodrigues}
\end{equation}
where $\widetilde{\bm{\psi}}$ denotes the skew-symmetric matrix associated with $\bm{\psi}$, and $\bm{I}$ is the $3\times3$ identity matrix. This Rodrigues parameterisation avoids the redundancy of quaternions for the local cross-sectional orientation while providing a singularity-free representation for rotations up to $2\pi$, which is more than sufficient for the wing deformations considered here.

The strain measures, force strain $\bm{\gamma}$ and moment strain $\bm{\kappa}$, are defined as
\begin{align}
  \bm{\gamma}(s,t) &= \bm{C}^{Ba}\left(\bm{r}_0' + \bm{u}'\right) - \bm{e}_1, \label{eq:forceStrain}\\
  \bm{\kappa}(s,t) &= \left(\bm{C}^{Ba}\right)'\bm{C}^{aB}\bm{e}_1 - \bm{\kappa}_0, \label{eq:momentStrain}
\end{align}
where primes denote derivatives with respect to $s$, $\bm{e}_1 = [1,0,0]^\top$ is the unit tangent vector of the undeformed beam, and $\bm{\kappa}_0$ is the initial curvature. The force strain $\bm{\gamma}$ captures axial extension and transverse shear, while the moment strain $\bm{\kappa}$ captures bending in two planes and twist. In the geometrically-exact formulation, these strain measures are objective, i.e., they are invariant under superimposed rigid-body motions, which is essential for the consistent coupling with flight dynamics.

The constitutive law relates generalised strains to sectional forces $\bm{F}$ and moments $\bm{M}$ through the cross-sectional stiffness matrix:
\begin{equation}
  \begin{Bmatrix} \bm{F} \\ \bm{M} \end{Bmatrix}
  = \begin{bmatrix} \bm{S}_{11} & \bm{S}_{12} \\ \bm{S}_{12}^\top & \bm{S}_{22} \end{bmatrix}
  \begin{Bmatrix} \bm{\gamma} \\ \bm{\kappa} \end{Bmatrix},
  \label{eq:constitutive}
\end{equation}
where the sub-matrices take the explicit forms
\begin{equation}
  \bm{S}_{11} = \begin{bmatrix} EA & 0 & 0 \\ 0 & GA_2 & 0 \\ 0 & 0 & GA_3 \end{bmatrix}, \quad
  \bm{S}_{22} = \begin{bmatrix} GJ & 0 & 0 \\ 0 & EI_2 & 0 \\ 0 & 0 & EI_3 \end{bmatrix}, \quad
  \bm{S}_{12} = \bm{0},
  \label{eq:stiffnessSubmatrices}
\end{equation}
for an isotropic, symmetric cross-section with no structural coupling. Here $EA$ is the extensional stiffness, $GA_2$ and $GA_3$ are the shear stiffnesses, $GJ$ is the torsional stiffness, and $EI_2$ and $EI_3$ are the bending stiffnesses about the two principal axes. For the HALE wing configuration studied here, $\bm{S}_{12} = \bm{0}$ (no bend-twist coupling), although the formulation accommodates non-zero coupling terms for composite or swept wings.

The equations of motion are obtained from Hamilton's principle. The virtual work of the internal and inertial forces leads to the weak form:
\begin{equation}
  \int_0^L \left[\delta\bm{\gamma}^\top \bm{F} + \delta\bm{\kappa}^\top \bm{M}\right] \mathrm{d}s
  + \int_0^L \left[\delta\bm{u}^\top \mu \ddot{\bm{u}} + \delta\bm{\psi}^\top \bm{J}_\rho \ddot{\bm{\psi}}\right] \mathrm{d}s
  = \int_0^L \delta\bm{u}^\top \bm{f}_{\mathrm{ext}} \,\mathrm{d}s,
  \label{eq:hamiltonPrinciple}
\end{equation}
where $\mu$ is the mass per unit length, $\bm{J}_\rho$ is the cross-sectional rotational inertia matrix, and $\bm{f}_{\mathrm{ext}}$ includes aerodynamic loads and gravity. The beam is discretised using two-noded finite elements with linear shape functions for the displacements and rotations. The semi-discrete system takes the form
\begin{equation}
  \bm{M}_s\,\ddot{\bm{\eta}} + \bm{C}_s(\bm{\eta},\dot{\bm{\eta}})\,\dot{\bm{\eta}} + \bm{K}_s(\bm{\eta})\,\bm{\eta} = \bm{f}_{\mathrm{ext}}(\bm{\eta},\dot{\bm{\eta}},t),
  \label{eq:structEOM}
\end{equation}
where $\bm{\eta}$ is the vector of nodal degrees of freedom, $\bm{M}_s$ is the consistent mass matrix, $\bm{C}_s$ is the velocity-dependent (gyroscopic and Coriolis) damping matrix, and $\bm{K}_s(\bm{\eta})$ is the tangent stiffness matrix. The dependence of $\bm{K}_s$ on $\bm{\eta}$ reflects the geometric nonlinearity: the stiffness changes with the current deformed configuration. The tangent stiffness matrix can be decomposed as
\begin{equation}
  \bm{K}_s(\bm{\eta}) = \bm{K}_e + \bm{K}_g(\bm{\eta}),
  \label{eq:tangentStiffness}
\end{equation}
where $\bm{K}_e$ is the linear elastic stiffness and $\bm{K}_g(\bm{\eta})$ is the geometric stiffness (also called the stress-stiffness or initial-stress stiffness), which accounts for the effect of the current internal forces on the structural stiffness. For upward-bending wings, the geometric stiffness introduces a tension-stiffening effect that increases the effective bending stiffness in the deformed configuration. This effect is absent from linear models and becomes significant at large tip deflections~\cite{hodges2003geometrically,palacios2010assessment}.

\subsection{Aerodynamic Model: Unsteady Strip Theory}
\label{subsec:aero}

The aerodynamic loads are computed using two-dimensional unsteady strip theory based on the Theodorsen formulation~\cite{theodorsen1935general,fung2008introduction}. Along each spanwise strip, the lift, moment, and drag are evaluated from the local effective angle of attack, which accounts for both the rigid-body motion and the elastic deformation of the wing. The strip theory approach, while less accurate than three-dimensional methods for low-aspect-ratio planforms, has been shown to provide adequate accuracy for HALE-type configurations with aspect ratios exceeding 20~\cite{patil2001nonlinear,daronch2014scitech}.

The total lift per unit span consists of circulatory and non-circulatory (apparent-mass) contributions:
\begin{equation}
  L(s,t) = L_{\mathrm{nc}}(s,t) + L_c(s,t).
  \label{eq:totalLift}
\end{equation}

The non-circulatory lift per unit span accounts for the inertia of the fluid displaced by the moving aerofoil:
\begin{equation}
  L_{\mathrm{nc}}(s,t) = \pi\rho b^2 \left[\ddot{h} + U\dot{\alpha} - b\,a\,\ddot{\alpha}\right],
  \label{eq:noncircLift}
\end{equation}
where $\rho$ is the air density, $b$ is the semi-chord, $U$ is the local freestream velocity, $\alpha$ is the effective angle of attack, $h$ is the plunge displacement, and $a$ is the elastic axis location (measured from mid-chord in semi-chords).

The circulatory lift per unit span is
\begin{equation}
  L_c(s,t) = 2\pi\rho U b\, C(k)\left[\dot{h} + U\alpha + b\left(\tfrac{1}{2}-a\right)\dot{\alpha}\right],
  \label{eq:theodorsenLift}
\end{equation}
where $C(k)$ is the Theodorsen function with reduced frequency $k = \omega b/U$. The aerodynamic moment per unit span about the elastic axis is
\begin{equation}
  M_{\mathrm{ea}}(s,t) = \pi\rho b^2\left[-b\,a\,\ddot{h} - U b\left(\tfrac{1}{2}-a\right)\dot{\alpha} - b^2\left(\tfrac{1}{8}+a^2\right)\ddot{\alpha}\right]
  + 2\pi\rho U b^2\left(a+\tfrac{1}{2}\right) C(k)\left[\dot{h} + U\alpha + b\left(\tfrac{1}{2}-a\right)\dot{\alpha}\right].
  \label{eq:theodorsenMoment}
\end{equation}

For time-domain simulations, the Theodorsen function is replaced by its time-domain equivalent via the Wagner function $\Phi(\tau)$~\cite{wagner1925,jones1938operational}:
\begin{equation}
  \Phi(\tau) = 1 - \psi_1 e^{-\varepsilon_1 \tau} - \psi_2 e^{-\varepsilon_2 \tau},
  \label{eq:wagner}
\end{equation}
where $\tau = Ut/b$ is the non-dimensional time, and $\psi_1 = 0.165$, $\psi_2 = 0.335$, $\varepsilon_1 = 0.0455$, $\varepsilon_2 = 0.3$ are the R.T.~Jones approximation coefficients~\cite{jones1938operational}. The circulatory lift in the time domain becomes a Duhamel convolution integral:
\begin{equation}
  L_c(s,t) = 2\pi\rho U b \int_0^t \Phi(t-\tau)\,\frac{\mathrm{d}w_{3/4}}{\mathrm{d}\tau}\,\mathrm{d}\tau,
  \label{eq:duhamel}
\end{equation}
where $w_{3/4}$ is the downwash at the three-quarter chord point. The Duhamel integral is evaluated using an augmented-state approach, introducing two additional aerodynamic states per strip to capture the lag dynamics:
\begin{align}
  \dot{x}_1(s,t) &= -\frac{\varepsilon_1 U}{b} x_1(s,t) + w_{3/4}(s,t), \label{eq:aeroState1}\\
  \dot{x}_2(s,t) &= -\frac{\varepsilon_2 U}{b} x_2(s,t) + w_{3/4}(s,t), \label{eq:aeroState2}
\end{align}
so that the circulatory lift can be written as
\begin{equation}
  L_c(s,t) = 2\pi\rho U b\left[(1-\psi_1-\psi_2)w_{3/4} + \frac{\varepsilon_1\psi_1 U}{b} x_1 + \frac{\varepsilon_2\psi_2 U}{b} x_2\right].
  \label{eq:augmentedLift}
\end{equation}
This augmented-state formulation avoids the need to store and evaluate the full convolution history, making it computationally efficient for long time-domain simulations.

For gust analysis, the K\"ussner function $\Psi(\tau)$ provides the indicial lift response to a sharp-edged gust~\cite{kussner1936zusammenfassender,leishman2006principles}:
\begin{equation}
  \Psi(\tau) = 1 - \phi_1 e^{-\beta_1 \tau} - \phi_2 e^{-\beta_2 \tau},
  \label{eq:kussner}
\end{equation}
with coefficients $\phi_1 = 0.5792$, $\phi_2 = 0.4208$, $\beta_1 = 0.1393$, $\beta_2 = 1.802$~\cite{leishman2006principles}. The gust-induced lift is computed using a Duhamel integral analogous to \eqref{eq:duhamel}, with the downwash replaced by the gust velocity component normal to the local chord. Two additional augmented states per strip are introduced for the gust dynamics, bringing the total number of aerodynamic states to four per strip.

The transformation between the aerodynamic frame $\mathcal{A}$ (aligned with the local freestream) and the beam cross-sectional frame $\mathcal{B}$ is achieved through the rotation matrix $\bm{R}_c$:
\begin{equation}
  \bm{v}^{(\mathcal{B})} = \bm{R}_c \, \bm{v}^{(\mathcal{A})},
  \label{eq:frameTransform}
\end{equation}
which accounts for the local sweep, dihedral, and twist angles induced by both the undeformed geometry and the elastic deformation. The effective freestream velocity at each strip is obtained by transforming the body-frame velocity through the quaternion-derived rotation matrix $\bm{R}_\zeta$:
\begin{equation}
  \bm{V}_{\mathrm{eff}}(s) = \bm{R}_\zeta^\top \bm{V}_\infty - \bm{v}_{\mathrm{body}}(s) - \dot{\bm{u}}(s),
  \label{eq:effectiveVelocity}
\end{equation}
where $\bm{V}_\infty$ is the inertial freestream, $\bm{v}_{\mathrm{body}}(s)$ is the velocity at station $s$ due to rigid-body motion, and $\dot{\bm{u}}(s)$ is the elastic velocity. The effective angle of attack at each strip is then
\begin{equation}
  \alpha_{\mathrm{eff}}(s) = \arctan\left(\frac{V_{\mathrm{eff},3}(s)}{V_{\mathrm{eff},1}(s)}\right),
  \label{eq:effectiveAoA}
\end{equation}
where $V_{\mathrm{eff},1}$ and $V_{\mathrm{eff},3}$ are the chordwise and normal components of the effective velocity, respectively. This expression naturally captures the contributions of rigid-body pitch, structural twist, and the geometric effect of wing bending (which rotates the local chord plane).

The aerodynamic drag is modelled using a parabolic polar:
\begin{equation}
  D(s,t) = \frac{1}{2}\rho V_{\mathrm{eff}}^2 c \left(C_{D_0} + \frac{C_L^2}{\pi e_0 AR}\right),
  \label{eq:drag}
\end{equation}
where $C_{D_0} = 0.01$ is the zero-lift drag coefficient and $e_0 = 0.95$ is the Oswald efficiency factor. For the high-aspect-ratio wings considered here, the induced drag contribution is small at the low lift coefficients typical of cruise, but becomes important at the higher angles of attack required for trim of very flexible configurations.

\subsection{Flight Dynamic Model: Quaternion-Based 6-DOF Equations}
\label{subsec:flightDyn}

The aircraft rigid-body dynamics are described by the Newton--Euler equations in the body-fixed frame. The translational equations are
\begin{equation}
  m\left(\dot{\bm{V}}_B + \bm{\omega}_B \times \bm{V}_B\right) = \bm{F}_{\mathrm{aero}} + \bm{F}_{\mathrm{grav}} + \bm{F}_{\mathrm{thrust}},
  \label{eq:translational}
\end{equation}
where $m$ is the total aircraft mass, $\bm{V}_B = [u, v, w]^\top$ is the body-frame velocity, $\bm{\omega}_B = [p, q, r]^\top$ is the angular velocity, and the right-hand side collects the aerodynamic, gravitational, and thrust forces resolved in the body frame. For a flexible aircraft, the aerodynamic force is obtained by integrating the distributed strip loads over the deformed wing:
\begin{equation}
  \bm{F}_{\mathrm{aero}} = \int_{-L}^{L} \bm{C}^{Ba}(s) \begin{bmatrix} -D(s) \\ 0 \\ L(s) \end{bmatrix} \mathrm{d}s,
  \label{eq:integratedAeroForce}
\end{equation}
where the rotation tensor $\bm{C}^{Ba}(s)$ transforms the local aerodynamic forces from the strip frame to the body frame, naturally accounting for the effect of wing deformation on the force direction.

The rotational equations are
\begin{equation}
  \bm{J}\,\dot{\bm{\omega}}_B + \bm{\omega}_B \times \bm{J}\,\bm{\omega}_B = \bm{M}_{\mathrm{aero}} + \bm{M}_{\mathrm{thrust}},
  \label{eq:rotational}
\end{equation}
where $\bm{J}$ is the inertia tensor of the aircraft. For flexible aircraft, the inertia tensor depends on the current deformed shape and is updated at each time step:
\begin{equation}
  \bm{J}(t) = \bm{J}_0 + \Delta\bm{J}(\bm{\eta}(t)),
  \label{eq:inertiaUpdate}
\end{equation}
where $\bm{J}_0$ is the undeformed inertia and $\Delta\bm{J}$ is the correction due to elastic deformation. The correction is computed from the displaced mass distribution:
\begin{equation}
  \Delta\bm{J}(\bm{\eta}) = \int_{-L}^{L} \mu(s)\left[\|\bm{u}(s)\|^2 \bm{I} - \bm{u}(s)\bm{u}(s)^\top\right] \mathrm{d}s
  + 2\int_{-L}^{L} \mu(s)\left[\bm{r}_0(s)^\top\bm{u}(s)\,\bm{I} - \mathrm{sym}\left(\bm{r}_0(s)\bm{u}(s)^\top\right)\right] \mathrm{d}s,
  \label{eq:inertiaCorrection}
\end{equation}
where $\mathrm{sym}(\cdot)$ denotes the symmetric part of a matrix. For the very flexible configurations considered here ($\sigma > 2$), the inertia correction $\Delta\bm{J}$ can be significant relative to the undeformed inertia, making its inclusion essential for accurate flight dynamic predictions.

The attitude of the aircraft is represented using unit quaternions $\bm{q} = [q_0, q_1, q_2, q_3]^\top$ with $\|\bm{q}\| = 1$, which avoid the gimbal lock singularity of Euler angles~\cite{diebel2006representing}. The quaternion kinematic equation is
\begin{equation}
  \dot{\bm{q}} = \frac{1}{2}\bm{\Omega}\,\bm{q}, \quad
  \bm{\Omega} = \begin{bmatrix}
    0 & -p & -q & -r \\
    p &  0 &  r & -q \\
    q & -r &  0 &  p \\
    r &  q & -p &  0
  \end{bmatrix}.
  \label{eq:quaternionKinematics}
\end{equation}

The rotation matrix from the inertial frame to the body frame is recovered from the quaternion as
\begin{equation}
  \bm{R}_\zeta = \begin{bmatrix}
    1-2(q_2^2+q_3^2) & 2(q_1 q_2 - q_0 q_3) & 2(q_1 q_3 + q_0 q_2) \\
    2(q_1 q_2 + q_0 q_3) & 1-2(q_1^2+q_3^2) & 2(q_2 q_3 - q_0 q_1) \\
    2(q_1 q_3 - q_0 q_2) & 2(q_2 q_3 + q_0 q_1) & 1-2(q_1^2+q_2^2)
  \end{bmatrix}.
  \label{eq:quaternionRotation}
\end{equation}

The position of the centre of gravity in the inertial frame is propagated by
\begin{equation}
  \dot{\bm{p}}_{\mathrm{cg}} = \bm{R}_\zeta^\top \bm{V}_B,
  \label{eq:cgPosition}
\end{equation}
providing the altitude and ground-track information needed to evaluate the gravitational force and to assess the trajectory response during gust encounters.

\subsection{Monolithic Coupling}
\label{subsec:coupling}

The three subsystems are coupled into a single monolithic system. Defining the total state vector $\bm{x} = [\bm{\eta}^\top, \dot{\bm{\eta}}^\top, \bm{x}_a^\top, \bm{V}_B^\top, \bm{\omega}_B^\top, \bm{q}^\top, \bm{p}_{\mathrm{cg}}^\top]^\top$, where $\bm{x}_a$ collects the augmented aerodynamic states, the coupled equations of motion are written as
\begin{equation}
  \bm{M}_T(\bm{x})\,\dot{\bm{x}} + \bm{C}_T(\bm{x})\,\bm{x} + \bm{K}_T(\bm{x}) = \bm{f}_T(\bm{x},t),
  \label{eq:coupledSystem}
\end{equation}
where $\bm{M}_T$, $\bm{C}_T$, and $\bm{K}_T$ are the coupled mass, damping, and stiffness matrices, respectively, and $\bm{f}_T$ is the total forcing vector. The coupling arises through three distinct mechanisms. The first is aerodynamic coupling, whereby the aerodynamic loads depend on both the elastic deformation (through the local angle of attack and the direction of the lift vector) and the rigid-body motion (through the freestream velocity and attitude). The second is inertial coupling, whereby the elastic deformation modifies the mass distribution and hence the inertia properties, affecting the rigid-body dynamics through the updated inertia tensor $\bm{J}(t)$. The third is kinematic coupling, whereby the elastic velocities contribute to the angular momentum of the aircraft, and the rigid-body rotation affects the effective boundary conditions of the structural model.

The monolithic formulation treats all coupling terms implicitly, avoiding the well-known stability issues associated with staggered (partitioned) coupling schemes, particularly at low mass ratios where the added-mass effect is significant~\cite{hesse2014reduced}.

The coupled system~\eqref{eq:coupledSystem} is integrated in time using an implicit Newmark-$\beta$ scheme ($\beta = 0.25$, $\gamma = 0.5$) with Newton--Raphson iteration at each time step for the nonlinear structural terms. The tangent matrix for the Newton--Raphson iteration is
\begin{equation}
  \bm{K}_T^{\mathrm{tan}} = \frac{\partial \bm{R}}{\partial \bm{x}_{n+1}} = \frac{1}{\beta\Delta t^2}\bm{M}_T + \frac{\gamma}{\beta\Delta t}\bm{C}_T + \bm{K}_T + \frac{\partial \bm{f}_T}{\partial \bm{x}},
  \label{eq:tangentMatrix}
\end{equation}
where $\bm{R}$ is the residual vector. Convergence is declared when $\|\bm{R}\|/\|\bm{f}_T\| < 10^{-8}$. The time step is chosen to resolve the highest structural frequency of interest, typically $\Delta t = 0.01$~s~\citep{tantaroudas2015scitech}, which provides adequate resolution of the structural modes of interest.

For the linearised stability analysis, the nonlinear system is linearised about a steady-state equilibrium $\bar{\bm{x}}$ to obtain the generalised eigenvalue problem:
\begin{equation}
  \left(\lambda^2 \bm{M}_T + \lambda \bm{C}_T + \bm{K}_T^{\mathrm{tan}}\right)\hat{\bm{x}} = \bm{0},
  \label{eq:eigenvalueProblem}
\end{equation}
where $\lambda = \sigma_r + i\omega$ are the complex eigenvalues and $\hat{\bm{x}}$ are the corresponding eigenvectors. The real part $\sigma_r$ determines the stability (negative = stable) and the imaginary part $\omega$ gives the frequency. This eigenvalue problem is solved using the implicitly-restarted Arnoldi method, targeting the eigenvalues nearest to the imaginary axis.

\section{Flexibility Parameter Study}
\label{sec:parameterStudy}

\subsection{Definition of the Non-Dimensional Stiffness Parameter}
\label{subsec:sigmaDefinition}

To systematically investigate the effects of structural flexibility on flight dynamics, we introduce a non-dimensional stiffness parameter $\sigma$ defined as
\begin{equation}
  \sigma = \frac{EI_{\mathrm{ref}}}{EI_{\mathrm{actual}}},
  \label{eq:sigmaDefinition}
\end{equation}
where $EI_{\mathrm{ref}}$ is a reference bending stiffness corresponding to a baseline HALE wing configuration, and $EI_{\mathrm{actual}}$ is the actual bending stiffness used in the analysis. A value of $\sigma = 0.001$ corresponds to a nearly rigid wing (actual stiffness 1000 times the reference), while $\sigma = 4$ represents a very flexible wing (actual stiffness one-quarter of the reference). The parameter $\sigma$ is applied uniformly to all entries of the stiffness matrix $\bm{S}$ in Eq.~\eqref{eq:constitutive}, so that the modified stiffness matrix becomes
\begin{equation}
  \bm{S}(\sigma) = \frac{1}{\sigma}\bm{S}_{\mathrm{ref}},
  \label{eq:scaledStiffness}
\end{equation}
ensuring that the ratios between bending, torsional, and extensional stiffnesses remain constant across the parametric study.

The physical interpretation of $\sigma$ can be related to the ratio of aerodynamic to elastic forces. For a uniform wing in steady flight, the tip deflection normalised by semi-span is approximately
\begin{equation}
  \frac{\delta_{\mathrm{tip}}}{L} \approx \frac{q_\infty c C_L L^3}{8\,EI_2} = \frac{\sigma\, q_\infty c C_L L^3}{8\,EI_{2,\mathrm{ref}}},
  \label{eq:tipDeflectionScaling}
\end{equation}
where $q_\infty = \frac{1}{2}\rho U^2$ is the dynamic pressure and $C_L$ is the wing lift coefficient. This scaling shows that $\sigma$ is directly proportional to the static tip deflection for a given flight condition, making it a natural parameter for characterising flexibility effects.

Crucially, the mass distribution is held constant across all values of $\sigma$. This isolates the effect of stiffness variation from that of mass changes, which would alter both the structural dynamics and the flight dynamics independently. In a real design, reducing stiffness would typically involve reducing structural mass as well, but the present parametric approach provides cleaner insight into the stiffness effect alone.

\subsection{Baseline Aircraft Configuration}
\label{subsec:baseline}

The baseline aircraft configuration is based on the HALE model of Patil et al.~\cite{patil2001nonlinear}, which has become a standard benchmark for very flexible aircraft analysis. The aircraft consists of a straight, unswept wing with a rigid fuselage at the mid-span. The properties are summarised in Table~\ref{tab:HALEproperties} and the geometry is shown in \Cref{fig:haleGeometry}.

\begin{table}[htbp]
  \centering
  \caption{Baseline HALE aircraft properties (after Patil et al.~\cite{patil2001nonlinear}).}
  \label{tab:HALEproperties}
  \begin{tabular}{@{}lll@{}}
    \toprule
    \textbf{Property} & \textbf{Value} & \textbf{Unit} \\
    \midrule
    Semi-span, $L$                     & 16          & m \\
    Total span                         & 32          & m \\
    Chord, $c$                         & 1           & m \\
    Aspect ratio, $AR$                 & 32          & -- \\
    Mass per unit length, $\mu$        & 0.75        & kg/m \\
    Bending stiffness, $EI_2$          & $2\times10^4$ & N\,m$^2$ \\
    Bending stiffness, $EI_3$          & $4\times10^6$ & N\,m$^2$ \\
    Torsional stiffness, $GJ$          & $1\times10^4$ & N\,m$^2$ \\
    Moment of inertia (torsion), $j_t$ & 0.1         & kg\,m \\
    Elastic axis location              & 50\% chord  & -- \\
    Centre of gravity location         & 50\% chord  & -- \\
    Fuselage mass                      & 50          & kg \\
    Payload                            & 0           & kg \\
    \bottomrule
  \end{tabular}
\end{table}

The flight conditions representative of HALE operations are: altitude $h = 20{,}000$~m, freestream velocity $U = 25$~m/s, and air density $\rho = 0.0889$~kg/m$^3$. The corresponding dynamic pressure is $q_\infty = 27.8$~Pa and the Reynolds number based on chord is $Re \approx 1.5 \times 10^5$.

The values of $\sigma$ investigated are: $\{0.001, 0.01, 0.1, 0.5, 1.0, 1.5, 2.0, 2.5, 3.0, 3.5, 4.0\}$. For each value, the following analyses are performed: nonlinear static aeroelastic equilibrium (trim), linearised stability analysis about the trimmed state (flutter and flight dynamic modes), and nonlinear dynamic response to a discrete ``1-minus-cosine'' gust.

\section{Results}
\label{sec:results}

\subsection{HALE Aircraft Validation}
\label{subsec:validation}

The coupled framework is first validated against published results for the HALE aircraft configuration described in Table~\ref{tab:HALEproperties}. The wing is discretised using a finite-element mesh of 100 nodes, yielding 1,200 structural degrees of freedom. The total system dimension, including rigid-body states, quaternions, and aerodynamic augmented states, is 2,016~\citep{tantaroudas2015scitech}.

\subsubsection{Static aeroelastic validation}

\begin{figure}[htbp]
  \centering
  \includegraphics[width=0.65\textwidth]{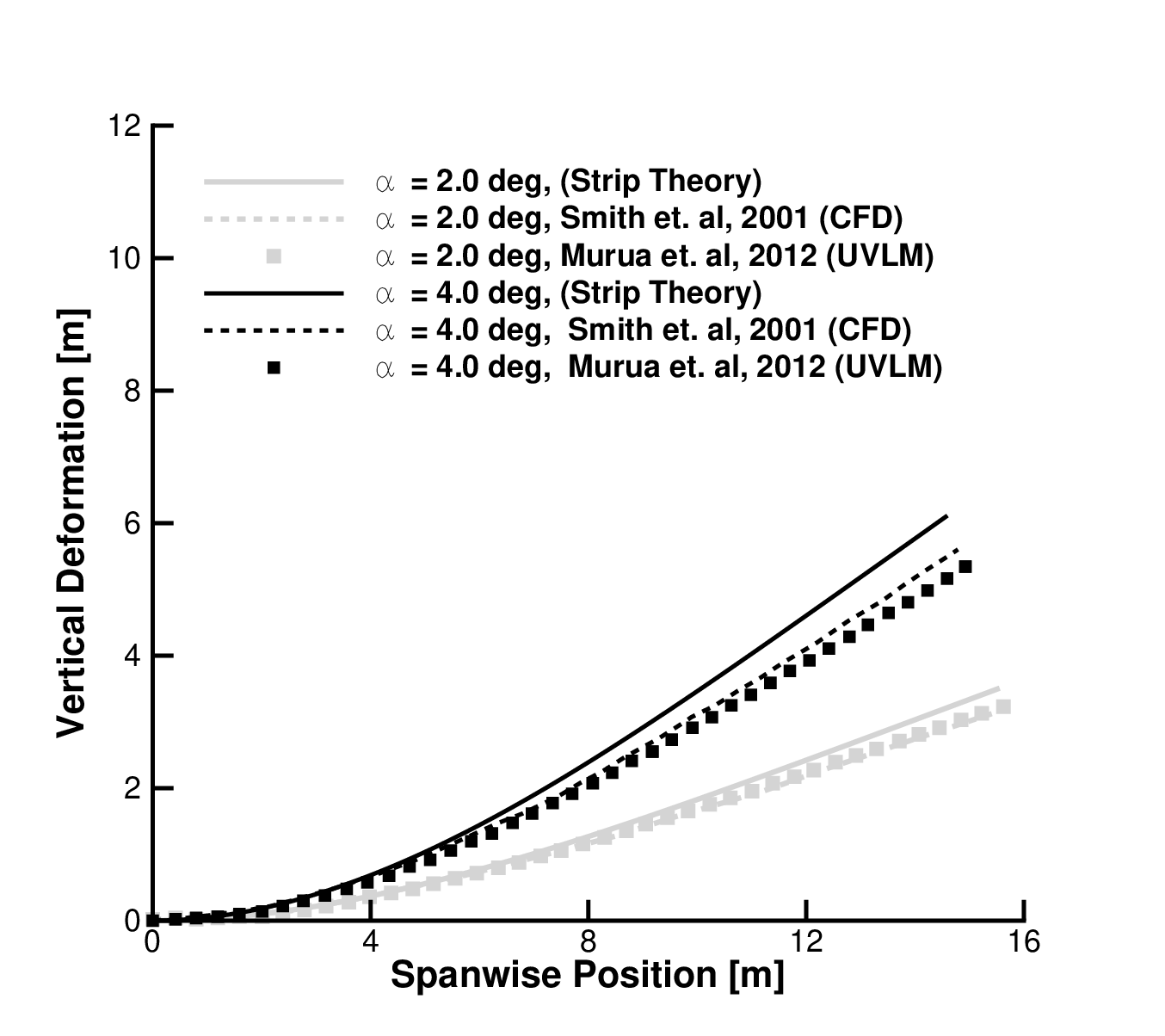}
  \caption{Static aeroelastic validation: vertical wing deformation along the semi-span at $\alpha = 2^\circ$ and $\alpha = 4^\circ$, comparing the present strip-theory model with CFD data and UVLM results ($U = 25$~m/s).}
  \label{fig:staticConvergence}
\end{figure}

The coupled framework is validated against published results by comparing the static wing deformation along the semi-span at two representative angles of attack. \Cref{fig:staticConvergence} compares the present strip-theory predictions with CFD data and UVLM results, demonstrating excellent agreement across the span~\citep{tantaroudas2015scitech}.

\subsubsection{Natural frequencies and mode shapes}

The natural frequencies of the uncoupled structural model are compared with published values in Table~\ref{tab:frequencies}. The present results are in excellent agreement with those of Patil and Hodges~\cite{patil2001nonlinear}, Murua et al.~\cite{murua2012applications}, and an exact beam theory solution.

\begin{table}[htbp]
  \centering
  \caption{Natural frequencies (rad/s) of the HALE wing: comparison with published data.}
  \label{tab:frequencies}
  \begin{tabular}{@{}lccc@{}}
    \toprule
    \textbf{Mode} & \textbf{Present} & \textbf{Patil~\cite{patil2001nonlinear}} & \textbf{Exact beam theory} \\
    \midrule
    1st bending & 2.24  & 2.24  & 2.24 \\
    2nd bending & 14.07 & 14.60 & 14.05 \\
    1st torsion & 31.04 & 31.14 & 31.04 \\
    1st in-plane bending & 31.71 & 31.73 & 31.71 \\
    3rd bending & 39.52 & 44.01 & 39.35 \\
    \bottomrule
  \end{tabular}
\end{table}

The structural frequencies of the first bending and torsional modes match well with previously published data~\cite{patil2001nonlinear}. The agreement confirms the accuracy of the finite-element discretisation.

\subsubsection{Static aeroelastic deflections}

\begin{figure}[htbp]
  \centering
  \includegraphics[width=0.65\textwidth]{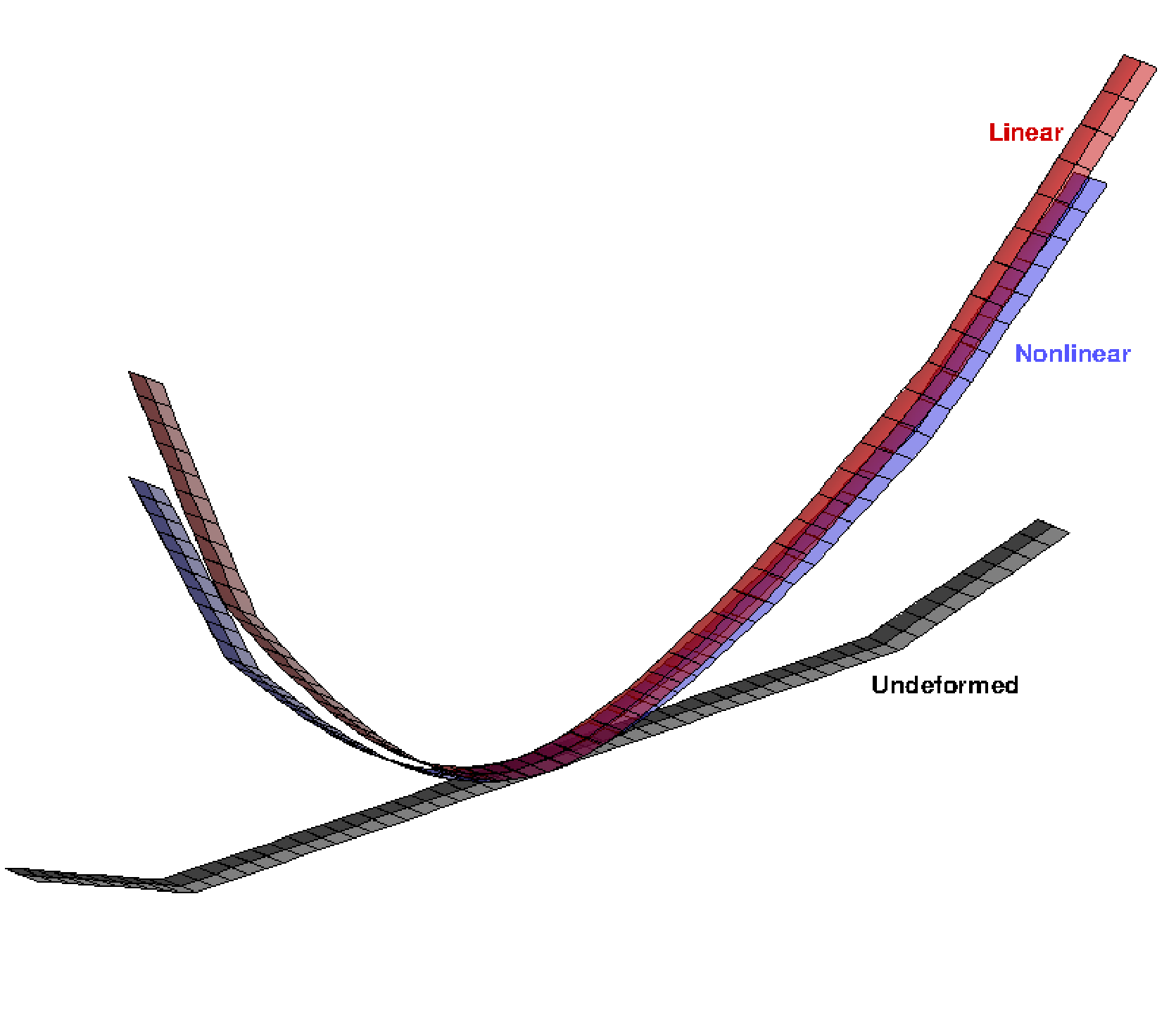}
  \caption{Three-dimensional deformed wing shapes: comparison of the linear analysis, the nonlinear geometrically-exact analysis, and the undeformed configuration ($\sigma = 1$, $U = 25$~m/s).}
  \label{fig:staticPatil}
\end{figure}

\Cref{fig:staticPatil} presents a three-dimensional comparison of the deformed wing shapes for the linear analysis, the nonlinear (geometrically-exact) analysis, and the undeformed baseline configuration.

The linear analysis overpredicts the tip deflection at higher angles of attack compared to the nonlinear solution. The geometric stiffening effect captured by the nonlinear formulation---arising from the tension-bending coupling represented by the geometric stiffness term $\bm{K}_g$ in \cref{eq:tangentStiffness}---is clearly visible in the reduced curvature of the nonlinear deformed shape. The progressive increase in effective dihedral with increasing load is also evident in \Cref{fig:staticPatil}.

\subsubsection{Flutter speed}

The flutter speed is determined via a $V$--$g$ analysis of the linearised coupled aeroelastic system about the undeformed equilibrium. The results are compared in Table~\ref{tab:flutter}.

\begin{table}[htbp]
  \centering
  \caption{Flutter speed (m/s) comparison for the HALE wing ($\sigma = 1$, undeformed equilibrium).}
  \label{tab:flutter}
  \begin{tabular}{@{}lc@{}}
    \toprule
    \textbf{Method} & \textbf{Flutter speed (m/s)} \\
    \midrule
    Present (strip theory)                        & 31.2 \\
    Patil and Hodges~\cite{patil2001nonlinear}     & 32.2 \\
    Murua et al.~\cite{murua2012applications}      & 33.0 \\
    \bottomrule
  \end{tabular}
\end{table}

The present flutter speed of 31.2~m/s is within 3\% of the Patil and Hodges value and within 6\% of the Murua et al.\ UVLM-based prediction. The slightly lower flutter speed from strip theory is consistent with its tendency to overpredict the unsteady aerodynamic loads in the absence of three-dimensional corrections. The flutter instability occurs through a coupling mechanism at a frequency of approximately 22~rad/s, involving the interaction of bending and torsion modes~\cite{patil2001nonlinear}.

\subsection{Static Wing Deformation Profiles}
\label{subsec:staticDeformation}

\begin{figure}[htbp]
  \centering
  \includegraphics[width=0.75\textwidth]{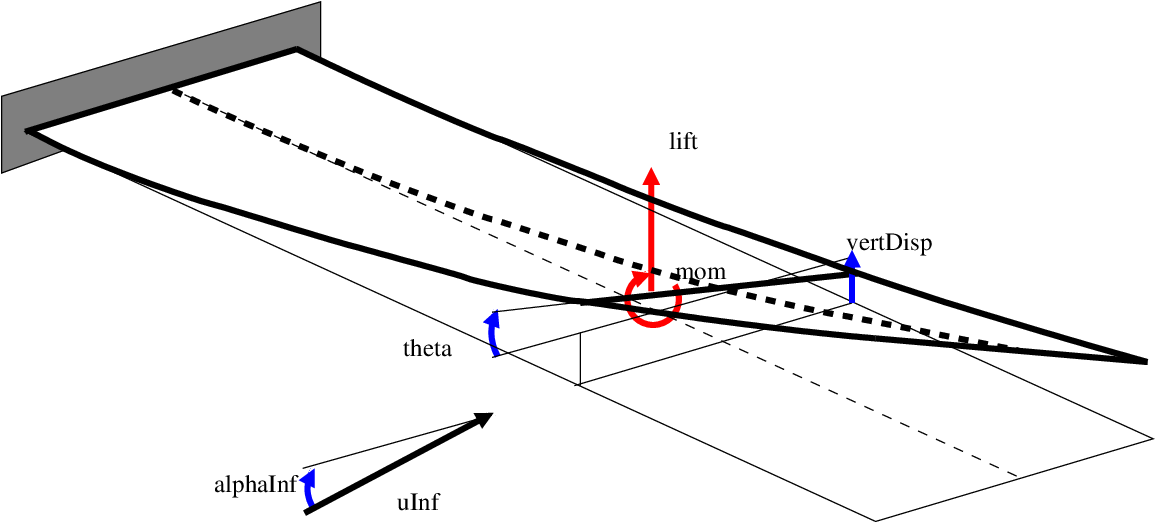}
  \caption{Schematic of the deformed wing showing the definitions of lift, pitching moment, vertical displacement, pitch angle ($\theta$), angle of attack ($\alpha_\infty$), and freestream velocity ($U_\infty$).}
  \label{fig:wingDeformation}
\end{figure}

Before proceeding to the full parametric study, we illustrate the key kinematic quantities that govern the coupled aeroelastic--flight dynamic behaviour. \Cref{fig:wingDeformation} presents a schematic of the deformed wing configuration, showing the definitions of lift force, pitching moment, vertical displacement, pitch angle, angle of attack, and freestream velocity used throughout the parametric study.

As the flexibility parameter $\sigma$ increases, the static wing deformation grows, and for low $\sigma$ values ($\sigma \leq 0.1$) the deformation closely follows the classical parabolic profile predicted by linear beam theory. For moderate flexibility ($\sigma = 0.5$--$1.0$), the deformation remains approximately parabolic but with larger amplitude. For high flexibility ($\sigma \geq 2.0$), the deformation profiles depart significantly from the parabolic shape: the curvature is concentrated near the root (where the bending moment is highest) and the outboard portion of the wing is relatively straight. This shape change is a direct consequence of the geometric nonlinearity---the tension induced by the large deflection stiffens the outboard portion of the wing, redistributing the curvature inboard.

At the highest flexibility levels, the wing-tip deflection reaches a large fraction of the semi-span, and the average geometric dihedral angle of the deformed wing becomes significant. This effective dihedral has profound consequences for the flight dynamics, as discussed in the following sections.

The evolution of the deformation profile with $\sigma$ also affects the spanwise load distribution. For rigid and moderately flexible configurations, the aerodynamic load distribution is approximately elliptical (assuming uniform chord and lift coefficient). For very flexible configurations, the inboard portion of the wing carries a disproportionate share of the load due to its more nearly horizontal orientation, while the outboard portion generates less vertical lift because its lift vector is rotated inward by the local dihedral angle. This load redistribution is an inherently nonlinear phenomenon that cannot be captured by superposition of linear mode shapes.

\subsection{Trim Analysis: Flexible vs.\ Rigid}
\label{subsec:trim}

The trim analysis seeks the equilibrium angle of attack $\alpha_{\mathrm{trim}}$ and thrust $T_{\mathrm{trim}}$ that satisfy vertical force equilibrium ($L = W$) and pitching moment equilibrium ($M_{\mathrm{cg}} = 0$) for a given flight speed and altitude. For the flexible aircraft, this is a nonlinear problem because the wing deformation depends on the aerodynamic loads, which in turn depend on the deformation. The trim solution is obtained iteratively using Newton's method, with the full nonlinear aeroelastic system converged at each iteration.

\begin{figure}[htbp]
  \centering
  \includegraphics[width=0.65\textwidth]{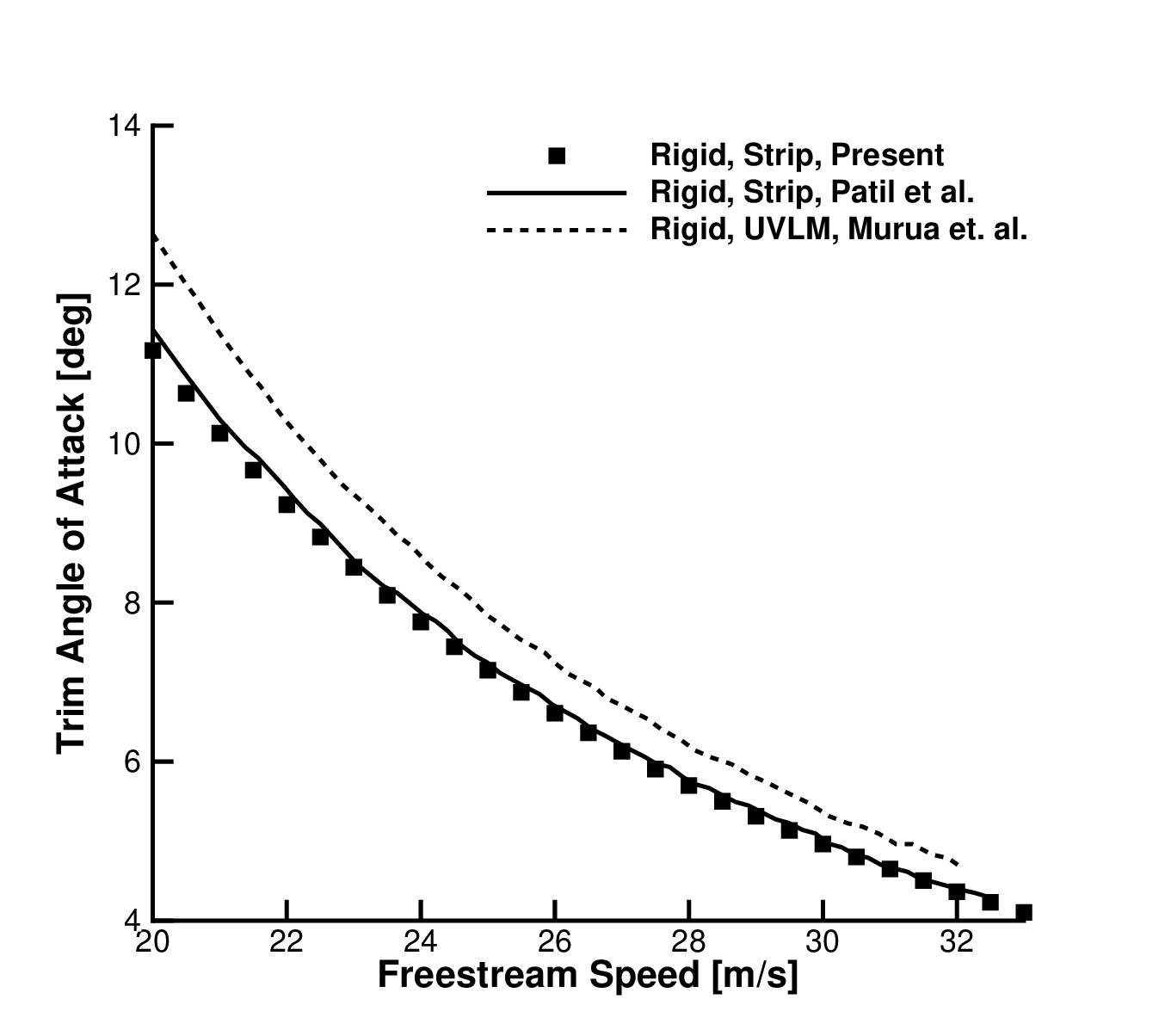}
  \caption{Trim angle of attack versus freestream speed for the rigid aircraft: comparison of the present strip-theory model (symbols) with Patil et al.~\cite{patil2001nonlinear} (solid line) and Murua et al.~\cite{murua2012applications} UVLM (dashed line).}
  \label{fig:trimAoA}
\end{figure}

\Cref{fig:trimAoA} shows the trim angle of attack as a function of freestream speed for the rigid aircraft, comparing the present strip-theory predictions with the results of Patil et al.~\cite{patil2001nonlinear} and the UVLM-based predictions of Murua et al.~\cite{murua2012applications}.

The present results are in good agreement with both reference datasets across the speed range. The trim angle of attack decreases with increasing speed as expected from the increase in dynamic pressure. Furthermore, a flexible wing requires a progressively larger angle of attack to achieve trim compared to a rigid wing due to the lift vector rotation effect described below.

The physical mechanism underlying this behaviour is the lift vector rotation effect. As the wing deforms upward under aerodynamic loading, the local lift vectors rotate inward, creating a net vertical lift component that is less than the total aerodynamic force magnitude. Consider a wing element at spanwise location $s$ with local dihedral angle $\Gamma(s)$ due to bending. The vertical component of the local lift is $L(s)\cos\Gamma(s)$, while the lateral (inboard) component is $L(s)\sin\Gamma(s)$. The total vertical force is
\begin{equation}
  F_z = \int_0^L L(s)\cos\Gamma(s)\,\mathrm{d}s < \int_0^L L(s)\,\mathrm{d}s,
  \label{eq:verticalForce}
\end{equation}
since $\cos\Gamma(s) < 1$ for any non-zero dihedral. In effect, the wing deformation creates a dihedral angle that redirects part of the lift force laterally rather than vertically. To compensate and maintain vertical equilibrium ($F_z = W$), the angle of attack must increase. This phenomenon was first identified by Patil et al.~\cite{patil2001nonlinear} and is confirmed quantitatively by the present results.

\begin{figure}[htbp]
  \centering
  \includegraphics[width=0.7\textwidth]{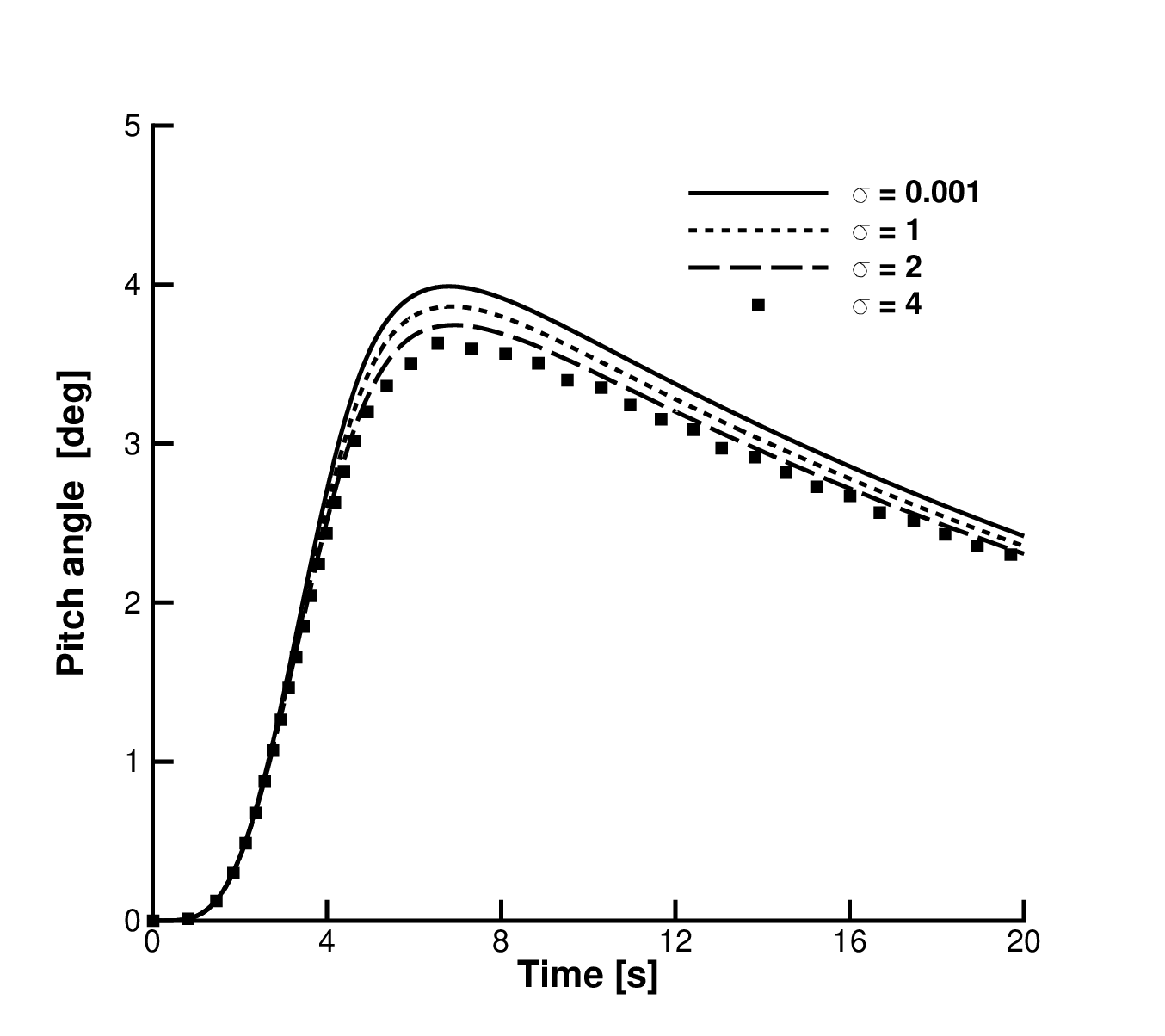}
  \caption{Pitch angle time history following a perturbation from trimmed equilibrium for $\sigma = 0.001$ (nearly rigid), $\sigma = 1$, $\sigma = 2$, and $\sigma = 4$ (very flexible). The nearly rigid case exhibits a well-damped response, while the very flexible case shows a lightly damped oscillation reflecting the coupling between structural and flight dynamic modes.}
  \label{fig:aoaFlexibility}
\end{figure}

The effect of flexibility on the pitch dynamics is shown in \Cref{fig:aoaFlexibility}, which presents the pitch angle time history following a perturbation from the trimmed equilibrium for different values of $\sigma$.

As the flexibility parameter $\sigma$ increases, the wing-tip deflection grows nonlinearly, reaching large fractions of the semi-span at the highest flexibility levels. The required trim angle of attack also increases with $\sigma$, as the effective lift-curve slope is reduced by the bending-induced dihedral. The rate of increase in $\alpha_{\mathrm{trim}}$ with $\sigma$ accelerates for higher flexibility levels, reflecting the increasing importance of the $\cos\Gamma$ factor as the effective dihedral grows.

\subsection{Trim and Flutter Speed Variation with Flexibility}
\label{subsec:flutterDegradation}

\begin{figure}[htbp]
  \centering
  \includegraphics[width=0.65\textwidth]{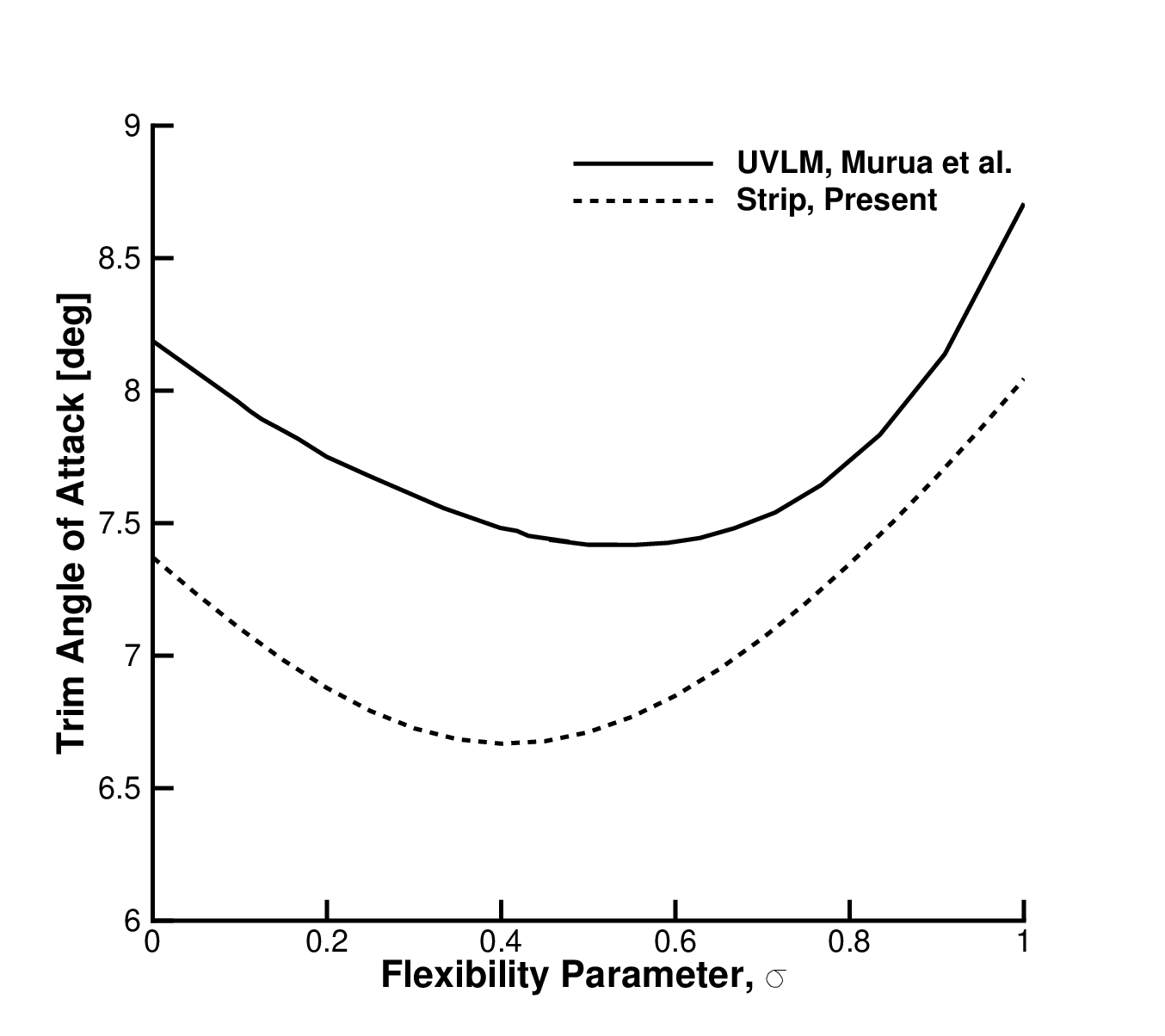}
  \caption{Trim angle of attack versus flexibility parameter $\sigma$: comparison between the present strip-theory model and the UVLM-based results of Murua et al.~\cite{murua2012applications}.}
  \label{fig:alphaFlutter}
\end{figure}

A critical question for the design of very flexible aircraft is how the trim conditions change with the level of structural flexibility. \Cref{fig:alphaFlutter} presents the trim angle of attack as a function of the flexibility parameter $\sigma$, comparing the present strip-theory predictions with the UVLM-based results of Murua et al.~\cite{murua2012applications}.

The trim angle of attack increases monotonically with $\sigma$, confirming the lift vector rotation mechanism discussed above. The two aerodynamic models show good agreement at low flexibility levels, with the strip-theory predictions diverging slightly at higher $\sigma$ values.

Regarding flutter speed degradation, when the flutter analysis is performed about the undeformed equilibrium, the flutter speed scales inversely with $\sqrt{\sigma}$ (since the bending stiffness scales as $1/\sigma$), giving $V_f \propto \sigma^{-1/2}$. This is the expected result from classical aeroelastic scaling.

However, when the pre-stressed equilibrium is accounted for, the flutter speed degradation is significantly more complex. For low $\sigma$ values, the two approaches give essentially identical results, as the static deformation is small and the pre-stress effects are negligible. For higher $\sigma$ values, the pre-stressed analysis predicts a higher flutter speed than the undeformed analysis, because the geometric stiffening effect increases the effective bending stiffness of the deformed wing.

This has important practical implications: conventional flutter analysis based on the undeformed geometry is conservative for flexible configurations, and accounting for the pre-stressed state can widen the flutter margin significantly. However, even with the pre-stress correction, the flutter speed still decreases monotonically with increasing $\sigma$.

The flutter mechanism also changes with flexibility. For low $\sigma$ values, the flutter instability involves the coalescence of the first bending and first torsion modes, as in the classical binary flutter problem. At higher flexibility levels, the flutter mechanism may change character as the modal spacing decreases with increasing flexibility.

\subsection{Flexibility Effect on Flight Dynamic Response}
\label{subsec:flightResponse}

The linearised stability analysis is performed about the trimmed equilibrium for each value of $\sigma$. The eigenvalues of the linearised system matrix yield the natural modes of the coupled aeroelastic--flight dynamic system. The flight dynamic modes of primary interest are the short-period and phugoid modes.

\begin{figure}[htbp]
  \centering
  \includegraphics[width=0.65\textwidth]{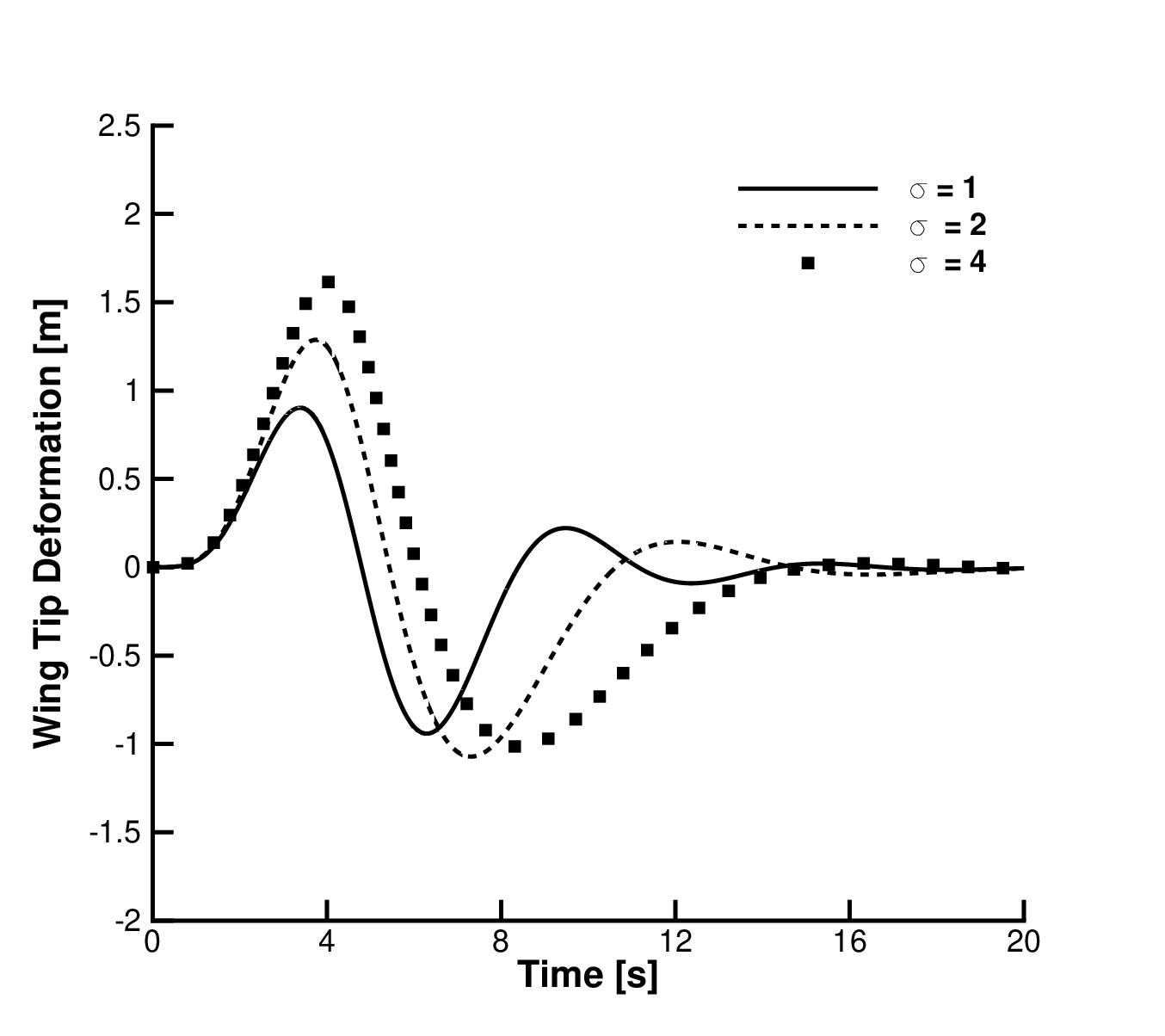}
  \caption{Wing-tip deformation time history during a gust encounter for $\sigma = 1$, $\sigma = 2$, and $\sigma = 4$. Higher flexibility levels produce larger oscillation amplitudes and reduced damping.}
  \label{fig:wtipFlexibility}
\end{figure}

\Cref{fig:wtipFlexibility} presents the wing-tip deformation time history during a gust encounter for three representative flexibility levels, illustrating the increasing oscillation amplitude and decreasing damping with higher $\sigma$.

\begin{figure}[htbp]
  \centering
  \includegraphics[width=0.65\textwidth]{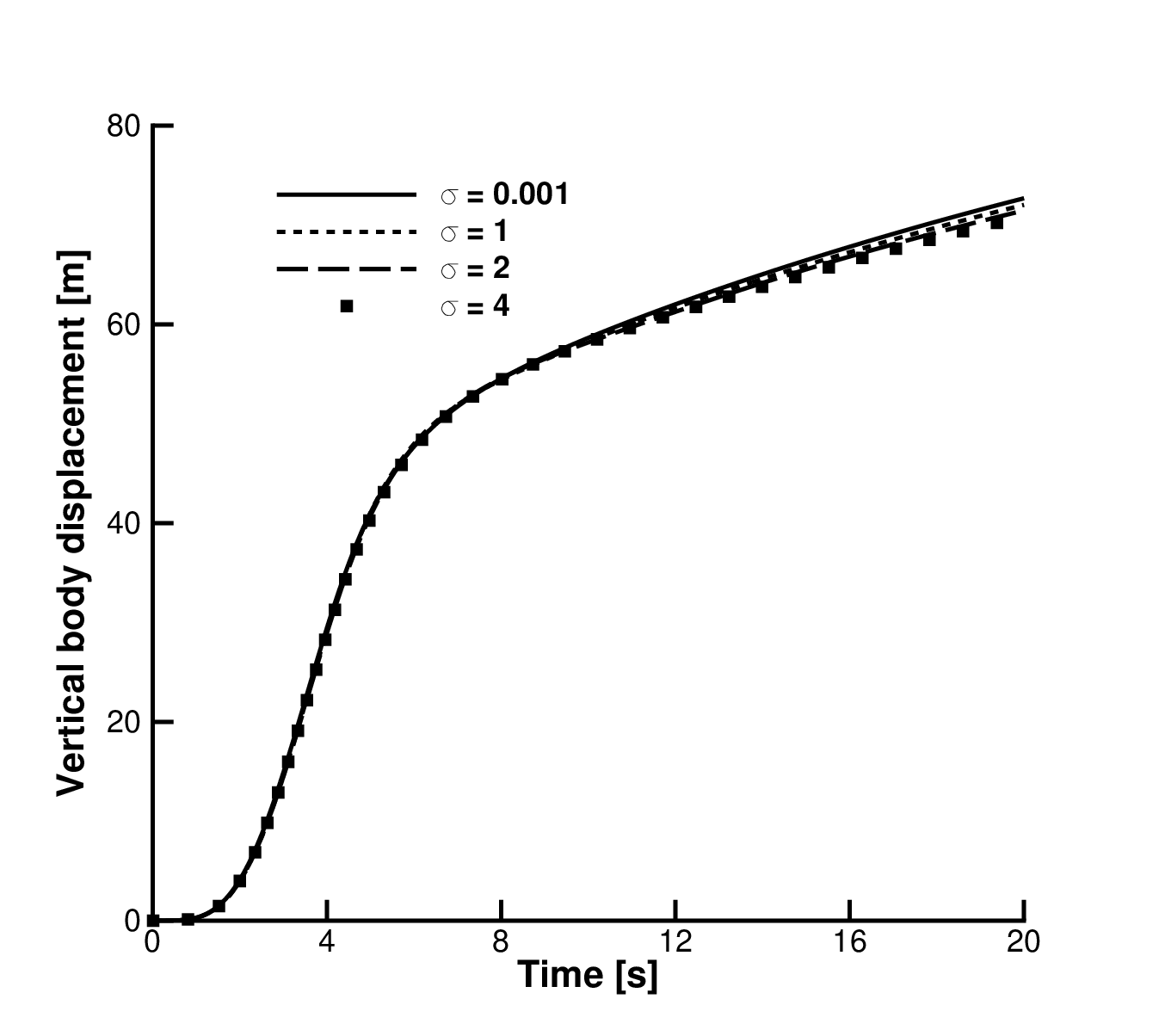}
  \caption{Centre-of-gravity vertical displacement time history for $\sigma = 0.001$, $\sigma = 1$, $\sigma = 2$, and $\sigma = 4$. The nearly rigid cases converge to a similar trajectory, while the very flexible case shows significant coupling with the structural dynamics.}
  \label{fig:vertdispFlexibility}
\end{figure}

\Cref{fig:vertdispFlexibility} shows the centre-of-gravity vertical displacement time history for different $\sigma$ values, demonstrating how the rigid-body trajectory response is increasingly coupled to the structural dynamics at higher flexibility levels.

The short-period mode remains stable throughout the range, with its damping and frequency decreasing modestly with increasing flexibility. The phugoid mode, however, shows a dramatic sensitivity to flexibility. At low $\sigma$ values, the phugoid is a lightly damped oscillatory mode consistent with classical rigid-body values. As $\sigma$ increases, the phugoid damping decreases monotonically and eventually becomes negative (unstable) at high flexibility levels~\cite{patil2001flight}.

This destabilisation of the phugoid mode by structural flexibility is one of the most significant findings. The physical mechanism involves two coupled effects. First, as the wing bends upward, the effective aerodynamic centre of the deformed wing shifts forward relative to the centre of gravity, weakening the pitch restoring moment that stabilises the phugoid. Second, the effective dihedral caused by bending reduces the lift-curve slope of the deformed wing (because only the $\cos\Gamma$ component of each strip's lift contributes to the vertical force), and this lower effective lift-curve slope reduces the pitch damping.

The large static deformation at high $\sigma$ values also creates an effective geometric dihedral angle $\Gamma_{\mathrm{eff}}$. This effective dihedral introduces lateral-directional coupling that is entirely absent in the rigid-wing model, demonstrating that longitudinal and lateral dynamics cannot be decoupled for very flexible configurations.

\subsection{Effect of Lift Vector Rotation on Flight Dynamics}
\label{subsec:liftVectorRotation}

The lift vector rotation mechanism deserves further quantitative analysis, as it is the primary driver of the flexibility-induced changes in trim and stability. For a wing element at spanwise location $s$ with local vertical displacement $u_3(s)$ and slope $\mathrm{d}u_3/\mathrm{d}s = \tan\Gamma(s)$, the lift vector in the inertial frame has components:
\begin{equation}
  \bm{L}_{\mathrm{inertial}}(s) = L(s) \begin{bmatrix} -\sin\alpha_{\mathrm{eff}} \\ -\sin\Gamma\cos\alpha_{\mathrm{eff}} \\ \cos\Gamma\cos\alpha_{\mathrm{eff}} \end{bmatrix},
  \label{eq:liftVectorInertial}
\end{equation}
where the first component is in the flight direction (contributing to induced drag), the second is lateral (creating a side force), and the third is vertical (supporting weight). The vertical force deficit, relative to a rigid wing, can be expressed as
\begin{equation}
  \Delta F_z = \int_0^L L(s)\left[1 - \cos\Gamma(s)\right]\mathrm{d}s \approx \int_0^L \frac{L(s)\Gamma^2(s)}{2}\,\mathrm{d}s,
  \label{eq:forceDeficit}
\end{equation}
where the approximation holds for moderate dihedral angles. Since the dihedral angle $\Gamma(s)$ increases with $\sigma$, the force deficit grows quadratically with the deformation level, requiring progressively higher trim angles of attack to maintain vertical force equilibrium.

The lateral force component created by the lift vector rotation is
\begin{equation}
  F_y = \int_0^L L(s)\sin\Gamma(s)\,\mathrm{d}s,
  \label{eq:lateralForce}
\end{equation}
which, for the symmetric configuration considered here, is zero in the trimmed state (the contributions from the left and right semi-spans cancel). However, during asymmetric perturbations (e.g., a lateral gust or a roll disturbance), the lateral force becomes non-zero and acts as a restoring force analogous to the dihedral effect in conventional aircraft. The effective dihedral derivative $C_{l_\beta}$ increases with $\sigma$, enhancing the Dutch roll stability but potentially causing excessive roll damping that could degrade manoeuvrability.

\subsection{Gust Response: Rigid vs.\ Flexible}
\label{subsec:gustResponse}

The dynamic gust response is investigated using a ``1-minus-cosine'' vertical gust profile:
\begin{equation}
  w_g(t) = \frac{w_{g0}}{2}\left(1 - \cos\frac{2\pi U t}{H_g}\right), \quad 0 \leq t \leq \frac{H_g}{U},
  \label{eq:gustProfile}
\end{equation}
where $w_{g0} = 5$~m/s is the peak gust velocity and $H_g = 25c = 25$~m is the gust gradient distance. The gust is applied uniformly across the span (i.e., simultaneous penetration) for the initial comparison. The gust velocity ratio $w_{g0}/U = 0.2$ represents a moderately severe gust encounter for HALE operations.

\begin{figure}[htbp]
  \centering
  \includegraphics[width=0.85\textwidth]{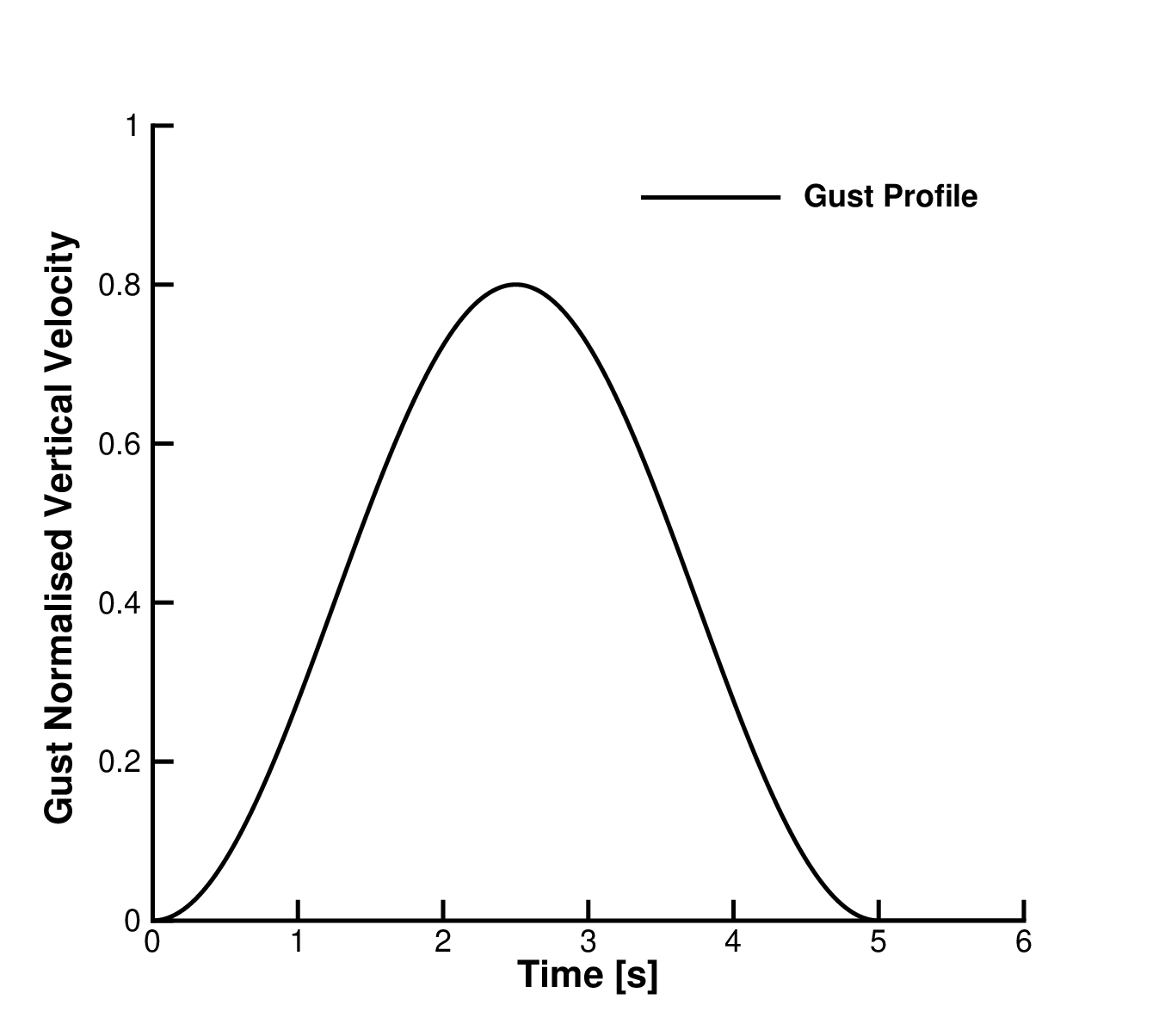}
  \caption{Normalised 1-minus-cosine gust velocity profile ($w_{g0} = 5$~m/s, $H_g = 25$~m). The gust-induced responses for different flexibility levels are shown in \Cref{fig:wtipFlexibility,fig:vertdispFlexibility}.}
  \label{fig:gustFlexibility}
\end{figure}

\Cref{fig:gustFlexibility} shows the normalised gust velocity profile used in the simulations.

For the nearly rigid configuration ($\sigma = 0.001$), the wing-tip deflection is negligible and the root bending moment shows a clean, quasi-steady response that closely follows the gust profile.

For the moderately flexible configuration ($\sigma = 1$), the wing-tip deflection increases significantly during the gust encounter. The root bending moment is reduced compared to the rigid case. This load alleviation is a well-known benefit of wing flexibility: the wing bends upward under the gust load, reducing the effective angle of attack and hence the peak aerodynamic force. After the gust, a lightly damped oscillation at the first bending frequency persists before decaying.

For the very flexible configuration ($\sigma = 4$), the response is qualitatively different. The wing undergoes large-amplitude oscillations that couple with the rigid-body pitch dynamics. The root bending moment shows a complex waveform with multiple frequency components. However, the peak local shear force at outboard span stations can be \emph{higher} for very flexible configurations, due to the inertial loads associated with the large wing-tip accelerations. This demonstrates that while flexibility generally provides root bending moment alleviation, it can redistribute loads in potentially adverse ways along the span.

A nonlinear geometric stiffening effect is observed at the highest flexibility levels. The wing deformation during the gust is so large that the effective structural stiffness increases (due to the tension-bending coupling in the geometrically-exact formulation), which limits the peak deflection below what a linear extrapolation would predict. This stiffening effect is absent from linearised models and represents an inherently nonlinear phenomenon that provides a natural limit on the maximum deformation.

For $\sigma = 4$, the coupling between the gust-induced wing oscillation and the rigid-body pitch dynamics excites the (now unstable) phugoid mode. Without active control, this oscillation grows in amplitude, confirming the stability analysis finding that the phugoid is unstable at high flexibility levels. The practical implication is that a gust encounter can trigger a divergent phugoid oscillation in very flexible aircraft, requiring the flight control system to actively suppress the response.

The load alleviation (reduction in peak root bending moment relative to the rigid case) increases monotonically with $\sigma$. However, the rate of alleviation decreases at high $\sigma$, following a diminishing-returns trend consistent with the geometric stiffening effect.

\section{Discussion}
\label{sec:discussion}

The parametric study reveals several findings of practical significance for the design of HALE and very flexible aircraft.

\paragraph{When is linear analysis acceptable?}
For low values of $\sigma$, corresponding to small wing-tip deflections, the differences between linear and nonlinear predictions are modest. In this regime, standard linear aeroelastic tools remain adequate for preliminary design, provided that the trim state is correctly computed. For higher $\sigma$ values, the nonlinear effects become significant and a geometrically-exact formulation is recommended~\cite{cesnik2012nonlinear}.

\paragraph{Phugoid destabilisation.}
The progressive destabilisation of the phugoid mode with increasing flexibility, confirmed in the present study, has important implications for flight control system design. At high flexibility levels, the open-loop phugoid is either marginally stable or unstable, and active control is required to maintain stable flight. The controller design must account for the coupling between structural and flight dynamic modes, as conventional decoupled autopilots may be inadequate~\cite{patil2001flight,cesnik2012nonlinear}. Active control strategies including $\mathcal{H}_\infty$ robust control and model reference adaptive control have been developed for gust load alleviation of such configurations~\cite{tantaroudas2014aviation,tantaroudas2015nonlinear}, and nonlinear controllers for flutter suppression have been demonstrated both in simulation and wind tunnel testing~\cite{daronch2014flutter,papatheou2013ifasd}. Reduced-order models suitable for real-time control implementation have been developed~\cite{tantaroudas2017bookchapter,tantaroudas2015scitech,daronch2013control}, addressing the computational cost barrier that would otherwise prevent model-based control of flexible aircraft.

\paragraph{Flutter speed degradation and the role of pre-stress.}
The flutter speed analysis reveals an important subtlety: the flutter boundary depends not only on the structural stiffness but also on the equilibrium state about which the stability analysis is performed. Conventional flutter analysis, based on the undeformed geometry, is conservative for flexible configurations because it neglects the geometric stiffening effect of the pre-stress. At high flexibility levels, the difference between the undeformed and pre-stressed flutter predictions becomes significant. This finding suggests that for very flexible aircraft, flutter clearance should be based on analysis about the correctly-computed nonlinear equilibrium, not the undeformed state. At higher flexibility levels, the flutter mechanism may change character as the modal spacing decreases, which may complicate flutter analysis based on tracking a single pair of eigenvalues.

\paragraph{Gust load alleviation vs.\ load redistribution.}
While flexibility provides root bending moment alleviation, the redistribution of loads along the span means that local structural elements outboard of the root may experience higher loads than in the rigid case. Structural sizing must therefore consider the full spanwise load distribution, not just the root bending moment. The geometric stiffening effect provides a natural protection against excessively large deformations, as the nonlinear analysis predicts lower peak tip deflections than the linear analysis at high flexibility levels. However, this stiffening is configuration-dependent and should not be relied upon without verification.

\paragraph{Strip theory limitations.}
The two-dimensional strip theory employed in this study provides adequate accuracy for the moderate angles of attack encountered in HALE cruise and gust conditions. However, for off-design conditions involving high angles of attack ($\alpha > 10^\circ$), stall, or significant spanwise flow, three-dimensional methods such as UVLM~\cite{murua2012applications} or CFD~\cite{wang2016nonlinear} are necessary. Strip theory can overpredict loads at high angles, which is non-conservative for flutter prediction but conservative for structural loads. The impact of aerodynamic modelling fidelity on manoeuvring aircraft has been systematically assessed in~\cite{daronch2014scitech}.

\paragraph{Comparison with higher-fidelity aerodynamics.}
The flutter speed predicted by the present strip theory (31.2~m/s) is approximately 3\% lower than the Patil--Hodges value and 6\% lower than the UVLM-based prediction of Murua et al. This systematic underprediction is consistent across configurations and can be accounted for through a correction factor in preliminary design. For detailed design and certification, coupled CFD--CSD analysis is recommended~\cite{wang2016nonlinear,deskos2020review}. The on-the-fly generation of aerodynamic tables from CFD~\cite{daronch2011onthefly} offers a promising intermediate-fidelity approach that captures three-dimensional and viscous effects while maintaining computational tractability for parametric studies.

\paragraph{Effective dihedral and lateral-directional coupling.}
The large static deformation at high flexibility ratios creates an effective dihedral that couples the longitudinal and lateral-directional dynamics. This coupling is not captured by conventional flight dynamic models that assume a symmetric, undeformed planform. For highly flexible configurations, a six-degree-of-freedom coupled analysis is essential, and the lateral-directional stability characteristics must be assessed alongside the longitudinal ones~\cite{hesse2016consistent,shearer2007nonlinear}. The effective dihedral derivative $C_{l_\beta}$ increases with flexibility, potentially causing issues with the Dutch roll mode and roll handling qualities. Active flutter suppression techniques~\cite{livne2018} may also need to account for the lateral-directional coupling in very flexible configurations.

\paragraph{Implications for HALE aircraft design.}
The results of this study provide concrete design guidance. For HALE aircraft operating at the flight conditions considered here ($U = 25$~m/s, $h = 20{,}000$~m), the following design thresholds emerge. At low flexibility levels, linear analysis is acceptable and conventional decoupled autopilots are sufficient. At moderate flexibility levels, nonlinear static analysis is recommended for trim computation, while linear stability analysis remains acceptable provided the correctly-computed equilibrium is used. At high flexibility levels, fully nonlinear coupled analysis is required, the phugoid becomes marginally stable or unstable, and active stability augmentation is recommended. At very high flexibility levels, the open-loop phugoid is unstable, active control is mandatory, the flutter margin is significantly reduced, and full 6-DOF coupled analysis is essential.

\section{Conclusions}
\label{sec:conclusions}

A coupled aeroelastic--flight dynamic framework has been developed and applied to a systematic parametric study of structural flexibility effects on the flight dynamics of high-aspect-ratio wings. The framework combines a geometrically-exact beam model, unsteady strip-theory aerodynamics with augmented-state lag terms, and quaternion-based 6-DOF flight dynamics in a monolithic coupling scheme. The main conclusions are as follows. Structural flexibility fundamentally alters the trim conditions, stability characteristics, and dynamic gust response of high-aspect-ratio wings; a flexible wing requires a larger angle of attack for trim due to the redirection of lift by wing deformation (the lift vector rotation effect), and the required thrust increases correspondingly due to the increased drag. The phugoid mode is progressively destabilised by increasing flexibility, eventually becoming unstable at high flexibility levels, while the short-period mode is relatively insensitive to flexibility; these findings are consistent with the results of Patil et al.~\cite{patil2001flight} and confirm that active flight control is essential for very flexible aircraft, with the physical mechanism involving both the forward shift of the effective aerodynamic centre and the reduction in effective lift-curve slope due to the bending-induced dihedral. Flutter speed degrades with increasing flexibility, but the rate of degradation depends critically on whether the analysis accounts for the pre-stressed equilibrium, as the geometric stiffening effect of the pre-stress increases the flutter speed relative to the undeformed analysis; at higher flexibility levels, the flutter mechanism may change character as the modal spacing decreases with increasing flexibility. The coupled framework reproduces published benchmark results with good accuracy: natural frequencies in close agreement with exact beam theory solutions, flutter speed within a few percent of established predictions~\cite{patil2001nonlinear,murua2012applications}, and static deflections in excellent agreement with the data of Patil and Hodges~\cite{patil2001nonlinear} at moderate angles of attack. Geometric stiffening provides a natural limit on deformation during gust encounters, reducing peak tip deflections compared to linear predictions; however, while flexibility provides root bending moment alleviation, outboard load redistribution can increase local structural loads, requiring consideration of the full spanwise load distribution. The stiffness parameter study provides guidance for design: linear analysis is acceptable at low flexibility levels, while a fully nonlinear, geometrically-exact formulation is required at higher flexibility levels, and at the highest flexibility levels, active flight control is mandatory due to phugoid instability. Strip theory is adequate for preliminary aeroelastic design at moderate angles of attack, but three-dimensional aerodynamic methods are recommended for detailed analysis, particularly at high angles of attack and for flutter speed prediction. Finally, fully coupled aeroelastic--flight dynamic analysis is essential for configurations with $\sigma > 1$, where the interaction between structural deformation, aerodynamic loads, and rigid-body dynamics produces qualitatively different behaviour from that predicted by decoupled approaches, and the effective dihedral created by large wing deformations introduces lateral-directional coupling that cannot be neglected.

Future work will extend the framework to include active control synthesis for phugoid stabilisation~\cite{tantaroudas2014aviation,tantaroudas2015scitech}, gradient-based stiffness optimisation, and coupling with higher-fidelity UVLM and RANS aerodynamics for transonic flexible aircraft applications. The development of reduced-order models~\cite{tantaroudas2017bookchapter,daronch2013control} for real-time control implementation will be pursued, as will the extension to asymmetric configurations (e.g., with control surface deflections or partial span damage). Experimental validation of nonlinear aeroelastic behaviour~\cite{fichera2014isma} using wind-tunnel models with calibrated flexibility levels will provide further confidence in the computational predictions.

\section*{Acknowledgements}

This work was supported by the U.K.\ Engineering and Physical Sciences Research Council (EPSRC) grant EP/I014594/1 on ``Nonlinear Flexibility Effects on Flight Dynamics and Control of Next-Generation Aircraft.'' The authors are grateful to Prof.\ A.\ Da Ronch and Prof.\ K.J.\ Badcock for their valuable guidance on flight dynamics modelling.
\bibliographystyle{unsrtnat}
\bibliography{references}

\end{document}